\documentclass[fleqn,usenatbib]{mnras}

\usepackage{newtxtext,newtxmath}
\usepackage{xcolor}

\usepackage[T1]{fontenc}

\DeclareRobustCommand{\VAN}[3]{#2}
\let\VANthebibliography\thebibliography
\def\thebibliography{\DeclareRobustCommand{\VAN}[3]{##3}\VANthebibliography}

\usepackage{graphicx}	
\usepackage{gensymb}
\usepackage{csquotes}

\title[N-rich Tidal Tails ]{APOGEE Detection of N-rich stars in the tidal tails of Palomar 5}

\author[S.~G.~Phillips et al.]{
Si{\^a}n G. Phillips$^{1}$,\thanks{E-mail: S.G.Phillips@2017.ljmu.ac.uk (SGP)}
Ricardo P. Schiavon$^{1}$,
J. Ted Mackereth$^{2,3}$, 
Carlos Allende Prieto$^{4,5}$,
Borja Anguiano$^{6}$,
\newauthor
Rachael L. Beaton$^{7,8}$,
Roger E. Cohen$^{9}$,
D. A. Garc{\'i}a-Hern{\'a}ndez$^{4,5}$,
Douglas Geisler$^{10,11,12}$,
Danny Horta$^{1}$,
\newauthor
Henrik J\"onsson$^{13}$,
Shobhit Kisku$^{1}$,
Richard R. Lane$^{14}$,
Steven R. Majewski$^{6}$,
Andrew Mason$^{1}$,
\newauthor
Dante Minniti$^{15, 16}$,
Mathias Schultheis$^{17}$,
and Dominic Taylor$^{1}$
\\
$^{1}$Astrophysics Research Institute, Liverpool John Moores University, 146 Brownlow Hill, Liverpool L3 5RF, UK\\
$^{2}$Canadian Institute for Theoretical Astrophysics, University of Toronto, 60 St. George Street, Toronto, ON, M5S 3H8, Canada
\\
$^{3}$Dunlap Institute for Astronomy and Astrophysics, University of Toronto, 50 St. George Street, Toronto, ON M5S 3H4, Canada \\
$^{4}$ Instituto de Astrof{\'i}sica de Canarias (IAC), E-38205 La Laguna, Tenerife,
Spain \\
$^{5}$ Universidad de La Laguna (ULL), Departamento de Astrof{\'i}sica, E-38206 La
Laguna, Tenerife, Spain \\
$^{6}$ Department of Astronomy, University of Virginia, Charlottesville, VA
22904-4325, USA \\
$^{7}$ The Observatories of the Carnegie Institution for Science, 813 Santa Barbara Street, Pasadena, CA 91101, USA \\
$^{8}$Department of Astrophysical Sciences, Princeton University, 4 Ivy Lane, Princeton, NJ, 08544, USA \\
$^{9}$ Space Telescope Science Institute, 3700 San Martin Drive, Baltimore, MD 21218 \\
$^{10}$ Departamento de Astronom{\'i}a, Universidad de Concepci{\'o}n, Casilla 160-C, Concepci{\'o}n, Chile \\
$^{11}$Instituto de Investigaci\'on Multidisciplinario en Ciencia y Tecnolog\'\i a, Universidad de La Serena. Avenida Ra\'ul Bitr\'an S/N, La Serena, Chile \\
$^{12}$ Departamento de Astronom\'\i a, Facultad de Ciencias, Universidad de La Serena. Av.
Juan Cisternas 1200, La Serena, Chile \\
$^{13}$ Materials Science and Applied Mathematics, Malm\"o University, SE-205 06 Malm\"o, Sweden\\
$^{14}$ Instituto de Astronom{\'i}a y Ciencias Planetarias de Atacama, Universidad de Atacama, Copayapu 485, Copiap{\'o}, Chile \\
$^{15}$ Departamento de Ciencias F{\'i}sicas, Universidad Andres Bello, Fern{\'a}ndez Concha 700, Las Condes, Santiago, Chile \\
$^{16}$ Vatican Observatory, V00120 Vatican City State, Italy \\
$^{17}$ University C\^ote d'Azur, Observatory of the C\^ote d'Azur, CNRS, Lagrange Laboratory, Observatory Bd, CS 34229, 06304 Nice cedex 4, France \\
}

\date{Accepted XXX. Received YYY; in original form ZZZ}

\pubyear{2021}

\begin{document}
\label{firstpage}
\pagerange{\pageref{firstpage}--\pageref{lastpage}}
\maketitle

\begin{abstract}
Recent results from chemical tagging studies using APOGEE data 
 suggest a strong link between the chemical abundance patterns of stars found within globular clusters, and chemically peculiar populations in the Galactic halo field. In this paper we analyse the chemical compositions of stars within the cluster body and tidal streams of Palomar 5, a globular cluster that is being tidally disrupted by interaction with the Galactic gravitational potential. We report the identification of nitrogen-rich (N-rich) stars both within and beyond the tidal radius of Palomar 5, with the latter being clearly aligned with the cluster tidal streams; this acts as  confirmation that N-rich stars are lost to the Galactic halo from globular clusters,
and provides support to the hypothesis that field N-rich stars identified by various groups have a globular cluster origin.
\end{abstract}




\section{Introduction}

Our understanding of the Galaxy's formation is greatly informed by identifying where stars in the Milky Way were born. To reconstruct the evolution of the Galaxy is to deduce a history of merger events, accretion and internal evolution; the study of these processes is especially  significant as hierarchical mass assembly is a keystone consequence of Lambda Cold Dark Matter ($\Lambda$-CDM) cosmology. Constraining the evolution of the Milky Way could therefore act as a test of $\Lambda$-CDM.  

\smallskip
A longstanding question is whether globular cluster (GC) dissolution has contributed to the stellar content of the Galaxy, and if so, to what extent \citep{Tremaine1975}.  Various studies over the past decade have identified field stars with abundance patterns characterised by enrichment in N and Na and depletion in C and O \citep[e.g.,][see \citet{ramirez2012} for the identification of the Na-O anticorrelation signature of GCs.]{martell2011building, lind2015gaia}. Such chemistry has long been associated with GC stars, suggesting that GC disruption may indeed contribute stars to the halo field.  More recently, a {\it weak} chemical tagging study by \cite{schiavon2017chemical} unveiled a large population of N-rich stars in the inner Galactic halo, which prompted further searches based on APOGEE data as well as other data sets \citep[e.g.,][]{ft2017,ft2019,tang2021} suggesting that the contribution of GC dissolution to the halo mass budget may be substantial \citep[see also, e.g.,][]{hughes2020alpha, horta2021contribution}. 

\smallskip The identification of putative GC escapees in the field is facilitated by the existence of chemical composition variations among stars in GCs. Although these chemical inhomogeneities have been documented long ago \citep[e.g.,][]{Kraft1979}, it was only after the detection of distinct evolutionary sequences in cluster CMDs \citep[e.g.,][]{Piotto2008}, and the measurement of detailed star-to-star abundance variations all the way down to the turnoff \citep[e.g.,][]{Cannon1998}, that the phenomenon was attributed to the presence of multiple stellar populations in GCs \citep[for reviews see e.g.,][]{Renzini2015,bastian2018multiple}.  In this framework,
 so-called ``first generation'' stars within GCs display chemical abundances indistinguishable from field stars at the same metallicity. Conversely, \enquote{second generation} GC stars are characterized by abundance patterns enhanced in elements such as He, N, Na, and Al, and diminished in C, O, and Mg \citep[for a review see e.g.,][]{bastian2018multiple}.  Because stars in the latter category can be discriminated from their field counterparts at the same metallicity, they are used as tracers of GC dissolution. Indeed, chemically peculiar halo stars identified in various studies occupy the same locus in the C-N, Na-O, and Al-Mg chemistry planes as ``second generation'' GC stars. 
 
 Early detections of N-rich stars in the halo suggested the contribution of GC disruption to the stellar mass budget to be at the few \% level \citep[e.g.,][]{martell2011building, martell2016chemical}. However, the identification of a large population in the inner halo by \cite{schiavon2017chemical} implied that the proportion of field stars which originate in GCs may be as high as $\sim$~25\% in the inner 2-3 kpc of the Galactic halo. This would imply that the mass contained within disrupted GCs would far exceed that locked in existing GCs. \cite{horta2021contribution}  used APOGEE DR16 data \citep{jonsson2020apogee} to measure the number density of field N-rich stars as a function of Galactocentric distance in the halo, placing the contribution from disrupted GCs to the field at $27.5^{+15.4}_{-11.5}$\% at r=1.5kpc, and $4.2^{+1.5}_{-1.3}$\% at r=10kpc, reconciling the measurements by \cite{martell2016chemical} and \cite{schiavon2017chemical}.
 
The results outlined above have important implications for our understanding of the formation and destruction of globular clusters \citep{hughes2020alpha} and the formation of the Galactic halo itself. Therefore, it is imperative to examine the underlying assumption that field N-rich stars and their GC counterparts share the same origin.


In this context, identifying N-rich stars as they are being shed by their parent GC would provide strong additional evidence in favour of the GC origin hypothesis for these types of field stars. 

The aim of this paper is to identify N-rich stars being lost from a GC currently undergoing disruption, and thus provide a snapshot of the journey of N-rich stars from the GC into the field. The natural starting point for such a search would be a GC with massive and well characterised tidal streams. While a significant proportion ($\sim 20\%$) of GCs exhibit tidal extensions \citep{kundu2019minniti}, Palomar 5 was chosen as the target for this study because of its well-defined tidal streams extending to over $\sim$30\degree on the sky \citep{starkman2020extended}. It is metal poor and has a low mass, with the majority of its initial mass believed to have now been lost to its streams, and has recently been identified as the host of a supramassive population of stellar mass black holes \citep{gieles2021}. For the purposes of this enquiry, its relevant features are the N-rich stars in the cluster centre and tidal streams.

\smallskip
The paper is structured as follows: in section \ref{sec:data} we introduce our data sources and the criteria by which our sample was selected. In section \ref{sec:results} we present and discuss the results, which are summarised in section \ref{sec:conclusions}.


\section{Data and Methods}
\label{sec:data}

This study makes use of stellar parameters and chemical abundances from the seventeenth data release of  SDSS-IV/APOGEE 2 \citep[][DR17 to be described in depth in Holtzmann et al. 2021, in prep]{blanton2017sloan, majewski2017apache}, and proper motions obtained from cross matching with the Gaia eDR3 catalogue \citep[][]{gaiamission, gaiaedr3} using TOPCAT software \citep{taylor2005topcat}. The public-source code \texttt{galpy}\footnote{(https://github.com/jobovy/galpy)} is used to create a model of radial velocities along the tidal streams \citep{GalpyCitation}.

\subsection{Data source}
\label{sec:dr17} 

The Apache Point Observatory Galactic Evolution Experiment (APOGEE 2) is a stellar spectroscopic survey within the Sloan Digital Sky Survey (SDSS-IV) which  has collected high quality NIR spectra for over 700,000 stars across all regions of the Galaxy. A detailed description of the target selection can be found in \cite{zasowski2017target, beaton2021target, santana2021target}.
APOGEE 2 operates from twin near-infrared (NIR) spectrographs \citep{wilson2019apache} installed on the 2.5m Sloan telescope at the Apache Point Observatory \citep{gunn20062}, and at the 2.5m du Pont telescope at Las Campanas Observatory \citep{bowen1973}, which essentially enables all-sky coverage. 

\smallskip

APOGEE 2 provides NIR spectra from which radial velocities, stellar parameters and detailed chemical compositions can be determined with high precision. The data from individual plates on individual observing nights are reduced using the APRED pipeline, with the APSTAR pipeline run following multiple individual visits, to resample the spectra obtained onto a fixed wavelength grid. The data finally enter the APOGEE Stellar Parameters and Chemical Abundances pipeline (ASPCAP) in which stellar parameters are analysed and calibrated, and abundances of up to 20 elements are determined. ASPCAP derives atmospheric parameters and abundances by matching synthetic spectra based on model
atmospheres \citep{gustafsson2008aspcap, meszaros2012aspcap} to the observations with the public code FERRE\footnote{https://github.com/callendeprieto/ferre} \citep{prieto2006aspcap}  ({for further details, see \citet{nidever2015data, perez2016aspcap,holtzman2015abundances,holtzman2018,jonsson2020apogee}}). 

\subsection{Sample selection}
The positions of the entire APOGEE DR17 sample were {\it initially} visualised using the TOPCAT software \citep{taylor2005topcat}  to identify those fields which were near-coincident with the streams as defined by a model calculated using {\tt galpy}; the data were limited to these fields of interest.
 
 
This {\tt galpy} model predicts the radial velocities and positions of stream members. The data from these fields of interest formed the parent sample, for which proper motions were obtained through a crossmatch with the Gaia eDR3 catalogue. The coordinates of the parent sample and radial velocity model were transformed from equatorial coordinates to $\phi$-space as defined by \cite{erkal2017sharper}, employing the transformation matrix supplied in their Appendix. The 
 {\tt galpy} models were found to agree approximately--to within a narrow angular range--with the 50th percentile model defined by \cite{erkal2017sharper} at  distances within $\sim$10\degree\  from the cluster 
centre, which was sufficient for this study. 

\bigskip

Candidate members of Palomar 5 were identified by selecting those stars that had a radial velocity measurement within 20kms$^{-1}$ of the stream model and were proximal in on-sky position. The tolerance in velocity difference was deliberately generous to account for model inaccuracies; we could only achieve an approximate fit to the position of the tidal streams with Galpy. This may be attributable to the simplicity of our Galpy model which does not account for aspects of dynamical history which may be specific to the streams under consideration, such as potential subhalo impacts and disc interactions. Further plausible sources of inaccuracies in the model include its sensitivities to the solar motion and Galactic potential; the potential implemented, \textbf{McMillan17}, may not be optimal for the case of Palomar 5. However, that a second selection method which did not use radial velocity as a criterion returned the same stellar sample (as described below) lends confidence to our ability to trust our Galpy model in the capacity to which we used it.

Globular cluster tidal radii are often uncertain, and in the case of Palomar 5 a range of measurements are found in literature: \cite{kuzma2015palomar} adopt $r_t~=~8.3$~arcminutes on the basis of a model by \cite{dehnen2004}, while \cite{xu2020new} determine $r_t = \sim 17.99 \pm 1.49$ arcminutes on the basis of a King model fit. The Gaia eDR3 update \citep{baumgardt2021} of the \cite{baumgardt2018catalogue} catalogue provides $r_t~=~7.2$~arcminutes. As the aim of this paper is the identification of stars beyond the tidal radius, we considered the largest of these values to avoid mistakenly identifying stars within the tidal radius as extra-tidal due to underestimating its extent. Palomar 5 members were considered to be those 
whose radial velocities differ from that of the cluster \citep[taken from][]{baumgardt2021} by no more than 10 km~s$^{-1}$. Noting that the velocity dispersion of Palomar 5 according to \cite{baumgardt2021} is only 0.6~km~s$^{-1}$, our adopted radial velocity tolerance was designed to maximise the candidate cluster member sample.

 \begin{figure}
\includegraphics[width=\columnwidth]{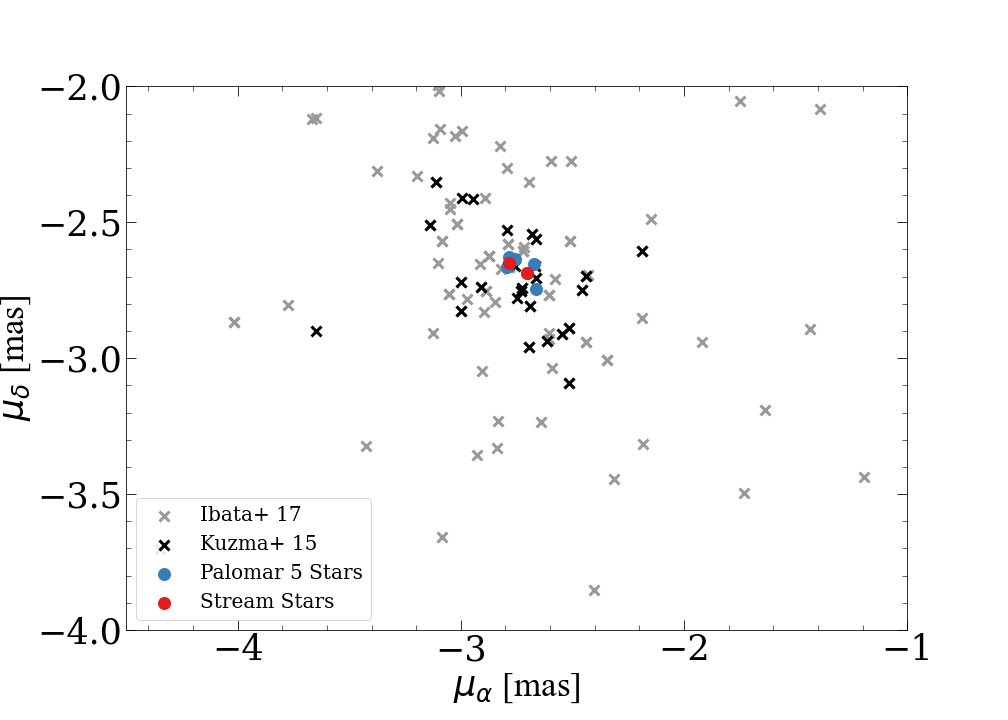}
\caption{Comparison of our sample stars in comparison with literature data in proper motion space. Stars shown in red and blue are those in the APOGEE DR17 catalogue meeting our membership criteria.} 
\label{fig:propermotions}
\end{figure}


The above sample of candidate stars was then restricted to those with combined spectra achieving a S/N$>$70 to ensure the reliability of their chemical composition measurements.   Additional cuts were made on certain physical parameters: a cut of log(g) < 3.6 was made to remove contamination by  foreground dwarf stars.  Effective temperature was limited to the range 3500 K < T$_{\textrm{eff}}$ < 4750~K to encompass all potential Palomar 5 stars while excluding both stars cool enough for their elemental abundances to be impacted by systematic effects, and stars hot enough to have unreliable abundances in C and N. The remaining sample was further proper-motion filtered through comparison with eDR3 data for Palomar 5 
 samples compiled by \citet{ibata2017feeling} and \citet{kuzma2015palomar}, as shown in Figure \ref{fig:propermotions}.  Stars with proper motions deviating by more than  2$\sigma$ of the mean proper motion of the Palomar 5 members identified by \citet{ibata2017feeling} were removed from the sample. None of those remaining candidates had an ASPCAP flag raised, indicating that their stellar parameters and chemical compositions are reliable.

Upon implementation of the aforementioned restrictions, the metallicity distribution of the surviving sample showed two peaks; one of these aligned with the approximate metallicity expected for Palomar 5, and the other was comprised of higher metallicity stars not associated with the cluster. Palomar 5 candidate members were then isolated by retaining only stars within the --1.5 < [Fe/H] < --1 interval. This left a sample of seven Palomar 5 candidate stars, which reflects the fact that quantity was a secondary consideration to purity in the final stage of our sample selection process. 

\smallskip
We checked our candidate sample by pursuing a second approach to selecting Palomar 5 stars in which no radial velocity criteria were adopted. We began by limiting the parent sample in temperature, gravity and SNR as before, and imposed spatial and metallicity cuts. The initial cut was made by limiting the parent data set to stars around the stream. We used a polynomial fit to the shape of the stream in $\phi$-space to select stars within one degree of the stream tracks defined by \cite{erkal2017sharper}. The remaining stars were distributed across a region in proper motion space defined by a small radius from the central values of proper motion for the Palomar 5 cluster, and thus no proper motion restriction was imposed. The metallicity distribution function of the stars formed two peaks, the secondary peak falling within the range $-1.5 <$ [Fe/H] $< -1$. The radial velocities of the stars forming this secondary peak were examined and those with radial velocities consistent with Palomar 5 formed the final sample. It was found to consist of the same stars as had comprised the sample obtained by the first selection method, adding confidence to the robustness of the stars' credibility as cluster members.

\begin{table}
	\centering
	\caption{APOGEE IDs of identified Palomar 5 stars}
	\label{tab:idtable}
	\begin{tabular}{@{}ccc@{}}
\hline
\hline
Star ID & Location & Type \\
\hline
2M15155284+0003219 & Cluster & N-rich \\
\hline
2M15155888-0005171 & Cluster & N-normal  \\
\hline
2M15160866-0008031 & Cluster & N-rich  \\
\hline
2M15162090-0008426 & Cluster & N-normal  \\
\hline
2M15163494-0017256 & Cluster &  N-normal \\
\hline
2M15183589+0027100 & Stream &  N-rich \\
\hline
2M15204588+0055032 & Stream & N-rich \\
\hline
\end{tabular}
\end{table}

\begin{figure}
	\includegraphics[width=\columnwidth]{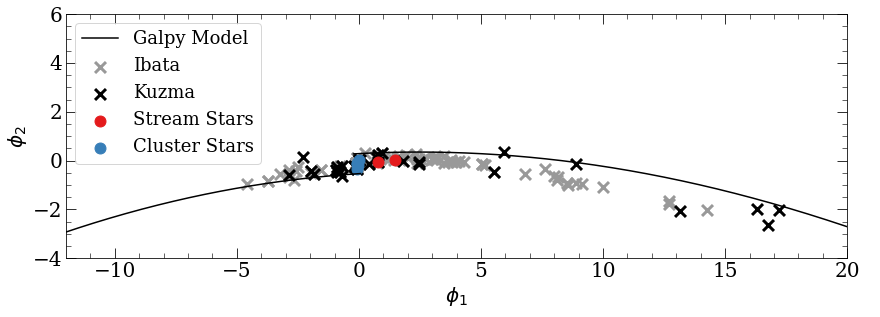}
    \caption{The sample of identified Palomar 5 stars is shown in comparison to the samples obtained from literature, distributed along the tidal stream models in $\phi$-space. Blue points are cluster stars, red points are extra-tidal stars aligned with the stream.}
    \label{fig:phispace}
\end{figure}

\begin{figure}
	\includegraphics[width=\columnwidth]{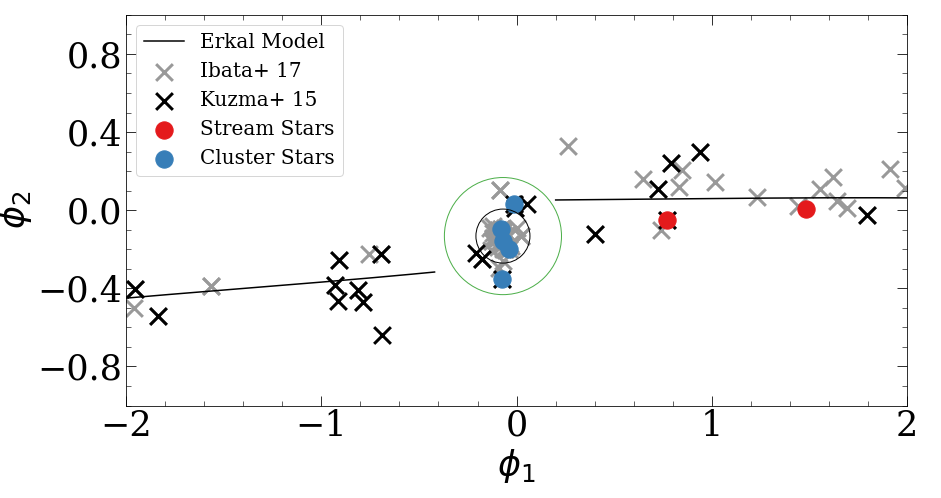}
    \caption{Literature samples of Palomar 5 stars are shown in comparison with the sample obtained by this study in $\phi$-space, focusing on the area immediately surrounding the cluster centre. Two values determined in separate works for the cluster tidal radius are shown: 8.3' from \citet{kuzma2015palomar} and 17.99' from \citet{xu2020new}. Stars we identify as members of Palomar 5 are indicated by red and blue points.  Two of our Palomar 5 stars are located beyond the cluster tidal radius and aligned with the trailing stream (to the right of the cluster on this plot).}
    \label{fig:central}
\end{figure}

The positions of the Palomar 5 stars thus identified are shown in comparison to literature samples along the extent of the stream in Figure \ref{fig:phispace} and in the vicinity of the cluster centre in Figure \ref{fig:central}. The IDs of these stars are listed in Table~\ref{tab:idtable}. 
In the next section, we examine the chemical properties of our sample of Palomar 5 members.

\section{Results}
\label{sec:results}

In this section, we study the distribution of Palomar 5 members in chemical composition planes, usually adopted for the diagnosis of multiple populations in GCs.  All the data employed in this analysis comes from APOGEE DR17.  
To illustrate the regions of chemical planes occupied by stars from different populations within GCs, the Palomar 5 sample was plotted alongside data for stars from the globular cluster M5 (NGC 5904).

This choice was made as M5 is a cluster which has been well-sampled by APOGEE and has a metallicity ([Fe/H] =--1.29) comparable to that of Palomar 5, so that its stars demarcate the locus of GC stars in various chemical planes, illustrating well-known anti-correlations.

In the following chemistry plots, elemental abundances and their associated errors are obtained from ASPCAP, where stellar parameters are derived from fitting the entire APOGEE spectrum. See \cite{jonsson2020apogee} for a detailed description of the determination of the elemental abundances and their errors.

\subsection{Carbon-Nitrogen}

\begin{figure}
	\includegraphics[width=\columnwidth]{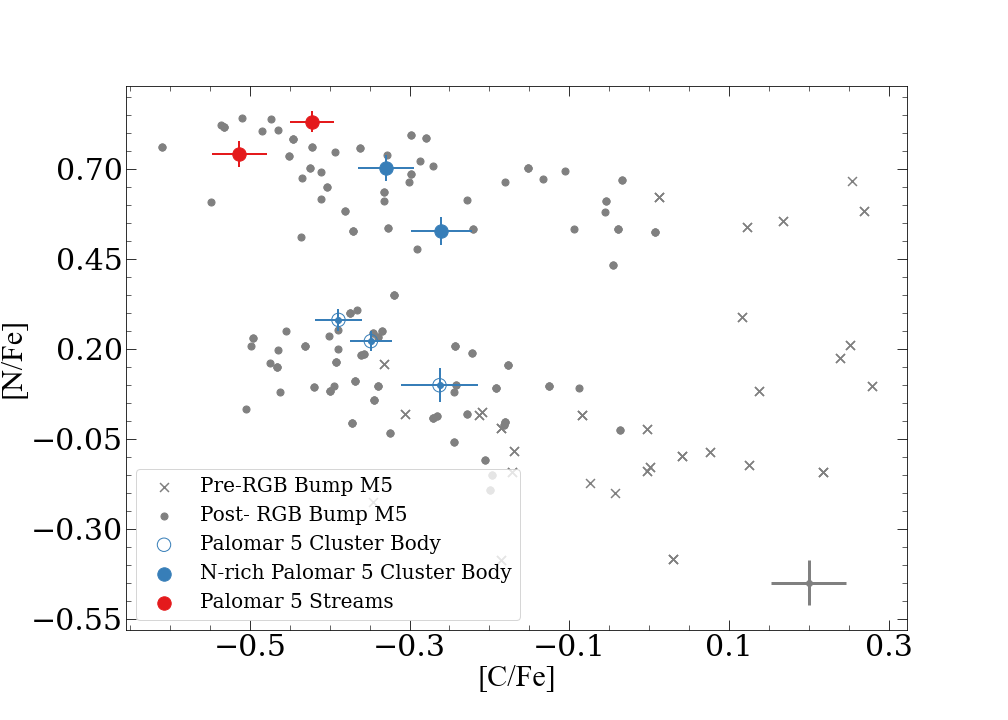}
    \caption{The Carbon-Nitrogen plane: this plot identifies members of the Palomar 5 tidal streams as N-rich. For the Palomar 5 stars, errorbars are plotted on data points, whereas for
M5 stars, a single mean errorbar is shown in the lower right of the plot.}
    \label{fig:cn}
\end{figure}

We first examine abundances in the carbon-nitrogen plane in Figure~\ref{fig:cn}, where stars are distributed along two distinct branches, N-normal and N-rich. The N-normal branch corresponds to the so-called \enquote{first generation} GC stars, which have chemical abundance patterns similar to field stars of the same metallicity. Conversely, the N-rich branch contains stars deemed to be so-called second generation. Each of the branches is characterised by an anticorrelation between the abundances of carbon and nitrogen, which is unrelated to the phenomenon of multiple populations known to characterise globular cluster populations \citep[e.g.,][]{Renzini2015, bastian2018multiple}. Instead, it is a consequence of the mixing that takes place during the evolution of low mass stars along the giant branch. 

 During the first dredge-up, a star's convection zone deepens as it evolves up the giant branch, reaching layers in which the abundances of carbon and nitrogen attained equilibrium levels through the CN(O) cycle during the star's main sequence lifetime \citep{salaris2020photometric}.  When the products of the CN(O) cycle are brought to the stellar surface the atmospheric abundances  of those elements are altered.  As stars evolve further up the giant branch, those abundances are further affected by additional non-convective mixing. The physical processes underlying this mixing are not well understood, with various potential explanations including stellar rotation and the influence of magnetic fields. The result is an increase of relative nitrogen abundance and a corresponding decrease in relative carbon abundance, hence the anticorrelation observed in C-N along both the N-rich and N-normal branches in Figure~\ref{fig:cn}. 
In order to bring this phenomenon into sharp relief, we adopt different symbols in Figure~\ref{fig:cn} for M5 stars above and below the Red Giant Branch (RGB) bump.

Now focusing on the Palomar 5 members, we see that they are distributed along both N-rich and N-normal branches, indicating the presence of the multiple populations phenomenon in this cluster. The two stars which were seen in the spatial plots to be aligned with the tidal streams and at distances beyond the tidal radius of the cluster are seen to be both enriched in nitrogen. One of these two stars has previously been identified as a likely extra-tidal star by \cite{trincado2019extratidal}. As seen in Figure~\ref{fig:hr} one of the N-rich stream stars appears to be on the RGB while the other aligns with the Asymptotic Giant Branch (AGB).

\begin{figure}
	\includegraphics[width=\columnwidth]{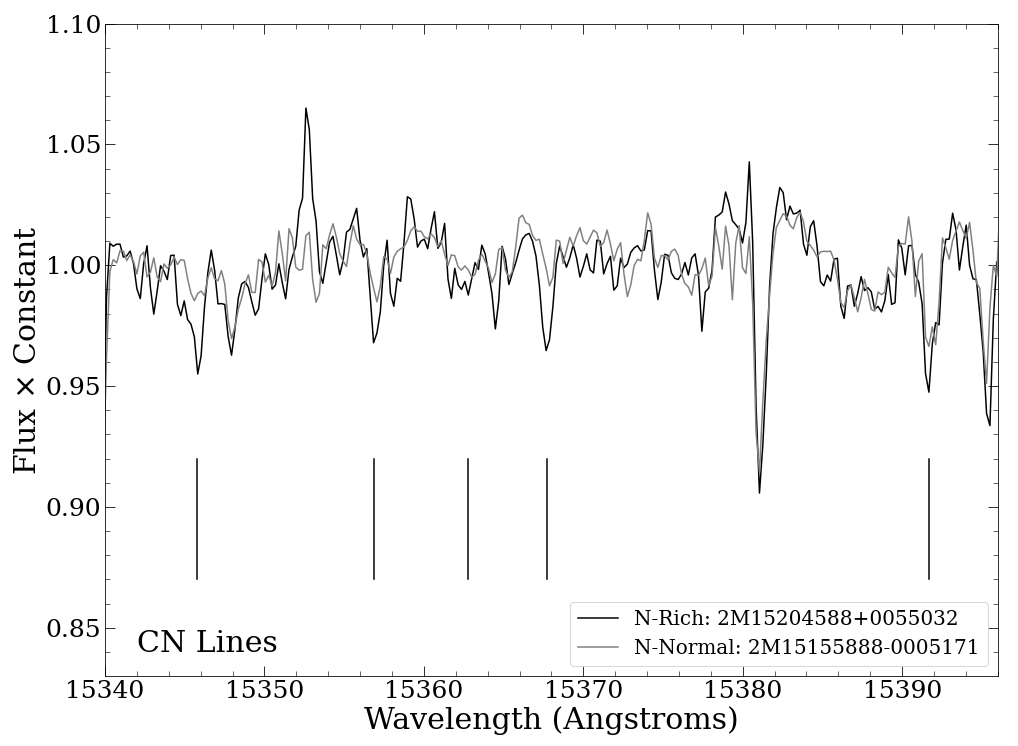}
    \caption{APOGEE spectra for an N-rich star (black line, APOGEE ID 2M15204588+0055032) and an N-normal star (grey line, APOGEE ID 2M15155888-0005171) where CN molecular lines are indicated by vertical markers.}
    \label{fig:cnlines}
\end{figure}

To further support the conclusion that the multiple populations phenomenon is present in our Palomar 5 sample, we show in Figure~\ref{fig:cnlines} a spectral comparison of a nitrogen rich and a nitrogen normal star which have otherwise nearly identical stellar parameters ($T_{\rm eff} \approx$ 4500K, log(g) $\approx 1.2$, [Fe/H] $\approx$ -1.3, [C/Fe] $\approx$ -0.5, [O/Fe] $\approx$ +0.3).

\medskip

\subsection{Magnesium-Aluminium}

\begin{figure}
	\includegraphics[width=\columnwidth]{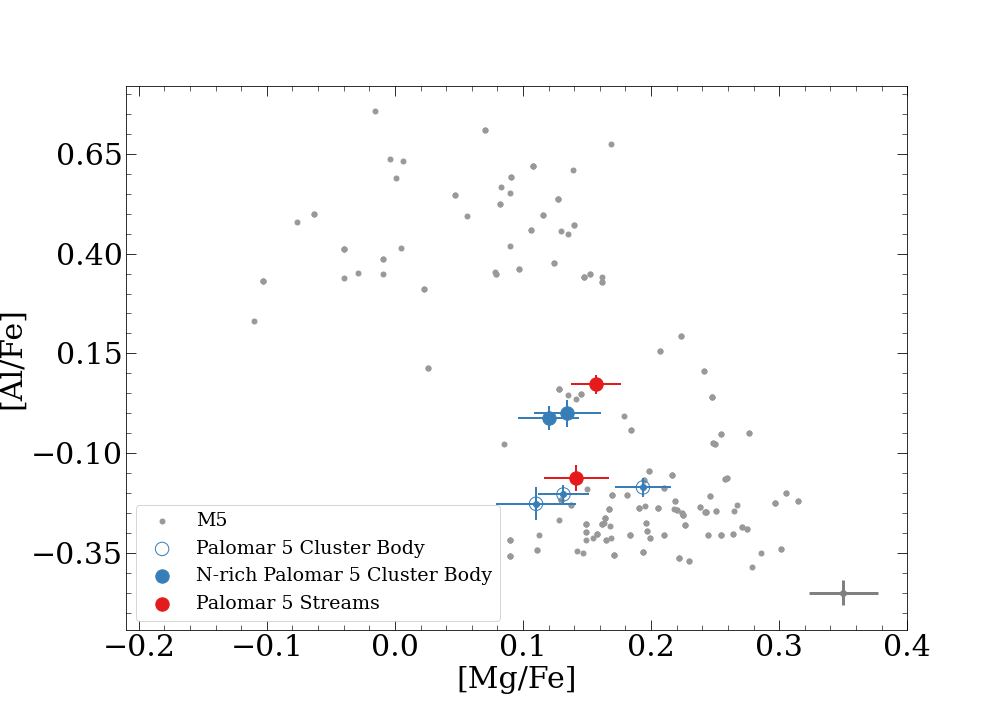}
    \caption{Magnesium-Aluminium plane: Palomar 5 stars do not show the anticorrelation observed for M5 stars, contrary to what would be expected for the cluster metallicity. Caution is required in the interpretation of this, given the small size of the sample.}
    \label{fig:mgal}
\end{figure}

 In Figure~\ref{fig:mgal} we display our sample stars in the Mg-Al plane.  An anticorrelation in Al-Mg can in some cases be a distinctive signature of ancient globular clusters, though it is not observed in all GCs which display anticorrelation in N-C and Na-O, and the extent of the Al-Mg anticorrelation is generally greater for clusters which are more massive and more metal poor \citep{carretta2009anticorrelation, ventura2016agb, pancino2017gaia}. As the Mg-Al cycle is initiated at T $\sim$ 50 MK, a temperature achieved in the core of a massive star only in the final stages of its main-sequence lifetime, various models have been suggested to account for the anticorrelation. These include pollution from asymptotic giant branch stars, a scenario from which the greater extent of the Al-Mg anticorrelation at lower cluster metallicities arises naturally \citep{ventura2016agb, bastian2018multiple}.

\smallskip

In the Mg-Al chemical plane, an anticorrelation is observed for M5. The Palomar 5 sample appears narrowly distributed, the range of Mg and Al abundances being far smaller than is observed in M5, which is unexpected for its metallicity. It is possible that the lack of a clear Mg-Al anti-correlation is a result of our small sample size.  Nevertheless, the ramifications of the Mg-Al anti-correlation being absent in Palomar 5 stars are worth considering.  

\cite{nataf2019relationship}, analysing a large cluster sample based on APOGEE DR14 \citep{dr14}, established a relation between GC metallicity and the minimum mass required for the presence of an Al-Mg anti-correlation:
\begin{equation}
\textrm{log}\left(\frac{\textrm{M}_\textrm{{GC}}}{\textrm{M}_{\odot}}\right) \approx 4.5 + 2.17([\text{Fe/H}] + 1.30).
\end{equation}
Using the value given in \cite{nataf2019relationship} of [Fe/H] = $-1.41$ for Palomar 5 in order to be consistent with the equation derived in their work, the  mass which would be required for Palomar 5 to display aluminium enrichment would be 1.83 $\times 10^4$ M$_{\odot}$. This is greater by an order of magnitude than the present-day cluster mass of $4.5-6 \times 10^3$ M$_{\odot}$ estimated by \cite{odenkirchen2002kinematic}, but below 3.96 $\times 10^4$ M$_{\odot}$, the lower limit imposed on the initial cluster mass prior to tidal dispersion by \cite{odenkirchen2003extended}. If the current cluster mass is the relevant parameter governing the onset of the Mg$-$Al anti-correlation, then its absence in Palomar 5 is to be expected. However, the multiple population phenomenon is observed in GCs spanning a wide age range \citep[e.g.,][]{Martocchia2018}, so it would naturally be expected that the {\it initial} GC mass is the determinant factor, and it is thus surprising that Palomar 5 does not display a Mg-Al anti-correlation.
For the sake of completeness, we point out that using the DR17 mean [Fe/H] value from our own sample of seven Palomar 5 stars, the minimum mass for aluminium enrichment would be 4.49 $\times$ $10^4$  M$_{\odot}$. This exceeds the lower limit on the aforementioned initial cluster mass estimation for Palomar 5. Nevertheless we consider it more appropriate to retain the value adopted by \cite{nataf2019relationship} for consistency with the DR14 metallicity scale.

The abundances for Palomar 5 suggest it has an accreted origin. Our sample of Palomar 5 stars has a mean [Si/Fe] = 0.132 and a mean [Fe/H] = $-1.23$, abundances which identify the cluster as accreted \citep[e.g.,][]{Danny2020GC, Shobhit}.  \cite{Massari2019} associate it with the Helmi streams on the basis of the cluster's integrals of motion.

\subsection{Nitrogen-Oxygen}

\begin{figure}
	\includegraphics[width=\columnwidth]{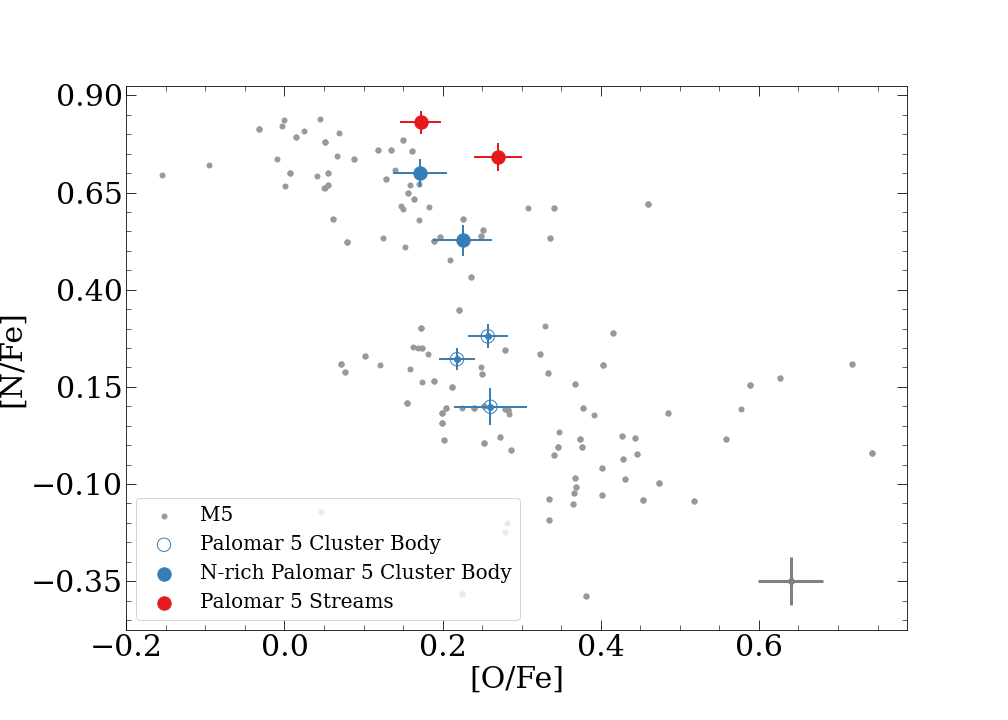}
    \caption{Nitrogen-Oxygen plane: Palomar 5 stars have a narrow range of oxygen abundances, following a distribution which is not consistent with that of M5.}
    \label{fig:no}
\end{figure}

\smallskip

In the N-O chemical plane (Figure~\ref{fig:no}), the Palomar 5 stars show a small spread in [O/Fe], which does not track the much wider variation seen in M5 stars. The narrow distribution in [O/Fe] may be a consequence of the small size of the stellar sample. \\


\subsection{Colour-Absolute Magnitude Diagram}

\begin{figure}
	\includegraphics[width=\columnwidth]{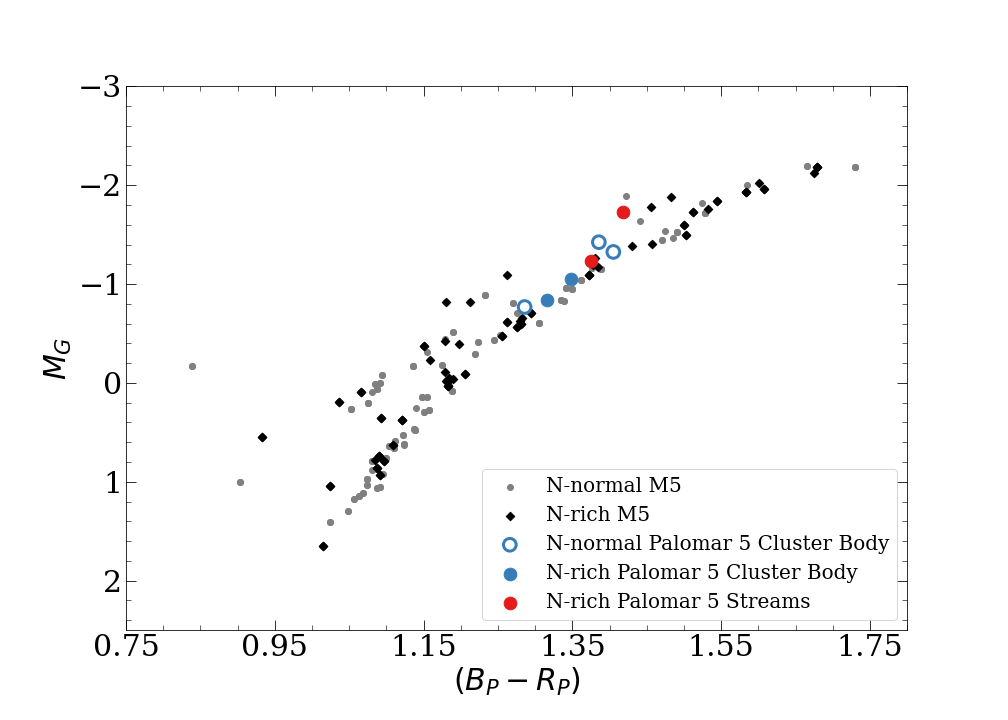}
    \caption{CMD using Gaia eDR3 data in which stars from Palomar 5 are seen to align with the evolutionary tracks of M5. One of the stream N-rich stars appears to be on the RGB, the other is likely an AGB star.}
    \label{fig:hr}
\end{figure}

In Figure~\ref{fig:hr} we display Palomar 5 and M5 stars in the Gaia eDR3 colour-absolute magnitude diagram to shed light on their evolutionary status. According to \cite{harris1996catalog} reddening is the same towards both clusters, at E(B$-$V) = 0.03; therefore, we do not correct the data for extinction. The data show N-rich and N-normal stars in M5 populating both the RGB and the AGB. Of the stars identified as belonging to Palomar 5, one of the stream stars appears aligned with the RGB, whereas the other appears aligned with the AGB.






\section{Conclusions}
\label{sec:conclusions}
Members of Palomar 5 were identified from APOGEE DR17 and Gaia EDR3. The selection criteria for members implemented cuts on position, radial velocity, effective temperature, surface gravity, metallicity, and S/N. This left a final sample of seven stars, which when plotted in chemistry planes, are found to be subdivisible into N-normal stars and N-rich stars, of which there are four. Two of the N-rich stars were positioned aligned with the trailing arm of the tidal stream, beyond the tidal radius of the cluster. This discovery is significant in providing unequivocal confirmation that N-rich stars born in globular clusters are lost to the field, and thus providing a bridge between both field and GC enriched populations.  Recent evidence to this association has been found by \cite{Trincado2021a}, who identified a chemically peculiar star just beyond the tidal radius of the bulge cluster NGC~6723. However, the optical photometry, stellar parameters, and high [C/Fe] of the star in question suggest that it may in fact be an N-normal AGB star.  Interestingly, \cite{Trincado2021b} identified the presence of N-rich candidates several tidal radii away from M~54, a globular cluster in the core of the Sagittarius dSph \citep[see][for an in depth discussion of M54 in the context of galaxy nucleation.] {bellazzini2008} The discovery by \cite{Trincado2021b} suggests that the process of field pollution by so-called ``second-generation'' cluster stars is taking place in the core of a dwarf galaxy undergoing tidal disruption during its merger with the Milky Way. It is important to note that tidal or Jacobi radii are notoriously uncertain, being subject to systematic errors related to their method of determination and the input parameters adopted. Thus, the more recent data from Gaia eDR3 provide a much greater value for the tidal radius than that which was used by \cite{Trincado2021b} which, considering the propagation of uncertainties of input parameters, would place the extra-tidal N-rich star discovered within 1$\sigma$ of the Jacobi radius of 37 arcminutes determined by \citet{vasilievEDR3baumgardt}.\footnote{https://people.smp.uq.edu.au/HolgerBaumgardt/globular/} We will present a discussion of tidal radius uncertainties in Schiavon et al. (2021, in prep.)


It is unexpected that the stars found to be enriched in nitrogen do not appear to be correspondingly enriched in aluminium. \cite{nataf2019relationship} propose a threshold current cluster mass at which aluminium enrichment would be expected to occur, which exceeds the present day cluster mass of Palomar 5. However, considering that most of the mass initially contained within the cluster has been lost to its streams, it may not be appropriate to consider the current mass but instead the initial mass, in which case the lack of observed aluminium enrichment in the \enquote{Second Generation} stars would be unexpected. To understand this phenomenon, a
more statistically significant sample of Palomar 5 stars is required. 

The association between the N-rich stars in GCs and their field counterparts relies entirely on the similarities of their chemical compositions, but there are alternative scenarios which could at least partially account for the presence of N-rich stars in the inner Galactic halo. It is possible that the field and GC enriched populations have evolved separately, and the greater proportion of enriched stars observed in GCs could be attributed, at least partially, to the relative abundance of data for GCs in comparison to the field. Other mechanisms may have contributed to abnormal enrichment of some field stars, such as AGB pollution. Understanding the extent of the contribution made by GC dissolution is a question which will benefit from the synergistic application of observational data and simulations, and from ever more extended and detailed chemodynamical information from surveys such as APOGEE. More in-depth analysis could be performed with a larger sample of stars obtained by targeting observations at the Palomar 5 streams, potentially in conjunction with extended stream models. This would allow for the investigation of more detailed questions, such as how the ratio of N-rich to N-normal stars in the cluster centre compares to that in the streams, and how this ratio changes as a function of distance along the stream. If a ratio of N-rich to N-normal stars could be established for the tidal streams of Palomar
5, and if it could be considered universally applicable to other globular clusters, this could
be used to infer the proportion of halo stars originating within GCs.

\smallskip

The discovery described in this paper, while providing additional support for the model of chemically peculiar field stars having their origin in globular clusters which have since been or are being disrupted, in no way represents a conclusive and complete validation of this explanation. The disparity between the metallicity distribution functions of field and GC N-rich stars must be considered in understanding the extent of the connection between these stellar populations. Additionally, the requirement for the mass of destroyed GC systems to greatly exceed the mass of existing GC systems is one which would need to be reconciled if this model were to be the sole or primary explanation for the origin of chemically peculiar field stars.

\section*{Acknowledgements}
The authors gratefully thank Julio Chaname, Eugene Vasiliev, Holger Baumgardt, Jos{\'e} Fern{\'a}ndez-Trincado, Cristina Chiappini, Adrian Price-Whelan and Allyson Sheffield for illuminating and fruitful discussion.
We thank the anonymous referee for their helpful and insightful comments on our manuscript. \\
The authors thank the many people whose hard work during these difficult pandemic times has made it possible for this research to have been carried out in safe conditions. 
DAGH acknowledges support from the State Research Agency (AEI) of the
Spanish Ministry of Science, Innovation and Universities (MCIU) and the European
Regional Development Fund (FEDER) under grant AYA2017-88254-P.
SRM acknowledges NSF grant AST-1909497.
DG gratefully acknowledges financial support from the Direcci\'on de Investigaci\'on y Desarollo de la Universidad de La Serena through the Programa de Incentivo a la Investigaci\'on de Acad\'emicos (PIA-DIDULS).

Funding for the Sloan Digital Sky Survey IV has been provided by the Alfred P. Sloan Foundation, the U.S. 
Department of Energy Office of Science, and the Participating Institutions. 

SDSS-IV acknowledges support and resources from the Center for High Performance Computing  at the University of Utah. The SDSS website is www.sdss.org.

SDSS-IV is managed by the Astrophysical Research Consortium for the Participating Institutions of the SDSS Collaboration including the Brazilian Participation Group, the Carnegie Institution for Science, Carnegie Mellon University, Center for Astrophysics | Harvard \& Smithsonian, the Chilean Participation Group, the French Participation Group, Instituto de Astrof\'isica de Canarias, The Johns Hopkins University, Kavli Institute for the Physics and Mathematics of the Universe (IPMU) / University of Tokyo, the Korean Participation Group, Lawrence Berkeley National Laboratory, Leibniz Institut f\"ur Astrophysik 
Potsdam (AIP),  Max-Planck-Institut f\"ur Astronomie (MPIA Heidelberg), Max-Planck-Institut f\"ur 
Astrophysik (MPA Garching), Max-Planck-Institut f\"ur Extraterrestrische Physik (MPE),
National Astronomical Observatories of China, New Mexico State University, New York University, University of 
Notre Dame, Observat\'orio Nacional / MCTI, The Ohio State University, Pennsylvania State 
University, Shanghai Astronomical Observatory, United Kingdom Participation Group, 
Universidad Nacional Aut\'onoma de M\'exico, University of Arizona, University of Colorado Boulder, 
University of Oxford, University of Portsmouth, University of Utah, University of Virginia, University
of Washington, University of Wisconsin, Vanderbilt University, and Yale University.

\section*{Data Availability}

This research draws on data from Gaia eDR3, which are publicly available, and from the 17th data release of the Sloan Digital Sky Survey (SDSS-IV), which are proprietary and will be released to the public in December 2021.



\bibliographystyle{mnras}
\bibliography{example} 








\bsp	
\label{lastpage}
\end{document}